\documentclass[12pt,a4paper]{article}
\usepackage[latin1]{inputenc}
\usepackage{amsmath}
\usepackage{amsfonts}
\usepackage{amssymb}
\usepackage{graphicx}
\usepackage{calc}
\usepackage{wrapfig}
\usepackage[a4paper,margin=1.0in]{geometry}
\usepackage{subfig}
\usepackage{cite}
\usepackage{hyperref}

\newcommand{\ie}{{\it{i.e.}}}
\newcommand{\eg}{{\it{e.g.}}}

\begin{document}
\title{The ILC positron target cooled by thermal radiation }
\author{S. Riemann$^1$\thanks{sabine.riemann@desy.de}, F. Dietrich$^1$, 
G. Moortgat-Pick$^2$, P. Sievers$^3$,  A. Ushakov$^2$\\
\\ $~$ 
 \normalsize $^1$\textit{Deutsches Elektronen-Synchrotron (DESY), Platanenallee 6, D-15738 Zeuthen}\\
 \normalsize $^2$\textit{University of Hamburg, Luruper Chaussee 149, D-22761 Hamburg} \\
 \normalsize $^3$\textit{CERN, CH-1211 Geneva 23, Switzerland}}

\maketitle
\begin{abstract}
\noindent
The  design of the
 conversion target for the undulator-based ILC positron source is still  under development. One important issue is the cooling of the target. Here, the  status of the design studies for cooling by thermal radiation is presented. 
\end{abstract}
  
\section{Introduction}\label{sec:intro}
The baseline design for the ILC positron source~\cite{ref:TDR} uses a long  helical undulator passed by the high energy electron beam to create  an intense circularly polarized  photon beam. The photon beam hits a thin conversion target to produce electron-positron pairs. The target is designed as wheel of 1\,m diameter and  spinning with 2000 rounds per minute in vacuum. This avoids overheating of the target material which is currently specified as Ti6Al4V. However,  water cooling of such positron conversion target is a challenge. Since the average energy  deposition in the target is only few kW, cooling by thermal radiation is a promising option which is currently studied. The status is presented here with focus on the $E_\mathrm{cm}=250\,$GeV option.

\section{The target parameters}\label{sec:params}
The ILC positron source is located at the end of the main linac. Since the photon yield and energy depend strongly on the electron energy, the source performance  has to be studied  for each centre-of-mass energy. The photon yield decreases with the electron energy and has to be compensated by a longer undulator. Due to the narrow photon beam the peak energy deposition density (PEDD) is high in spite of the spinning target. Both  PEDD as well as the average power deposited in the target vary for different $E_\mathrm{cm}$. Table~\ref{tab:sourcepar} summarizes the parameters  to achieve $Y= 1.5\,$e$^+/$e$^-$ as presented at POSIPOL 2016~\cite{ref:posipol16-AU}. Further studies~\cite{ref:AU-250GeVthickness} showed that for  $E_\mathrm{cm}=250\,$GeV a  reduced target thickness does not reduce the positron yield but decreases substantially the energy deposition in the target. Hence, for the 250\,GeV option the target thickness is 7\,mm. 
It turned out that for the 250\,GeV option the energy deposition in the inner part of the flux concentrator front side is too high. To resolve this problem, the drift space between undulator and target is reduced. Alternatively, a quarter wave transformer could be used. Further details can be found in reference~\cite{ref:AU-OMD}. Definitely, a distance of 401\,m between middle of the undulator and target represents the case with highest load. Thus, this option is studied and the results are presented in this paper.    
\begin{table}[h]
\begin{center}
\renewcommand{\baselinestretch}{1.2}
\begin{tabular}{|lc|cc|ccc|}
\hline
electron beam energy    & GeV & 126.5 & 125 & 150 & 175 & 250\\ \hline
undulator active length &  m  & \multicolumn{2}{c|}{231} &\multicolumn{3}{c|}{147}\\ 
undulator K             &     & \multicolumn{2}{c|}{0.85}& 0.8 & 0.66 & 0.45 \\
photon yield per m undulator & $\gamma$/(e$^- \, $m) & \multicolumn{2}{c|}{1.70} &  1.52 &1.07 & 0.52\\
photon yield            & $\gamma$/e$^-$     & \multicolumn{2}{c|}{392.7}& 223.9 & 157.3 & 76.1 \\ 
photon energy (1$^\mathrm{st}$ harmonic) & MeV & 7.7  & 7.5  & 11.3 & 17.6 & 42.9 \\
average photon energy                  & MeV & 7.5  & 7.3  & 10.4 & 13.7 & 26.8 \\
average photon beam power              &  kW & 62.6 & 60.2 & 48.8 & 45.2 & 42.9 \\
photon bunch energy                    &   J &  9.6 & 9.2 & 7.4 & 6.9 & 6.5 \\
electron energy loss in undulator     &  GeV &  3.0 & 2.9 & 2.3 & 2.2.& 2.0 \\
Ti6Al4V target thickness                & mm &  7   & 14.8 & \multicolumn{3}{c|}{14.8}\\
energy deposition per photon in target & MeV & 0.23 &  0.7 &  0.8 &  1.0 &  1.4\\
relative energy deposition              & \% &  3.1 &  9.0 &  8.0 &  7.3 &  5.3 \\
average power deposited in target       & kW &  1.94 &  5.4 &  3.9 &  3.3 &  2.3 \\
energy deposition per bunch             &  J &  0.3 & 0.83 & 0.60 & 0.50 & 0.35\\
space from middle of undulator to target & m &  401 & 570 & \multicolumn{3}{c|}{500}\\
photon beam spot size on target ($\sigma$) & mm &  1.2 & 1.72 & 1.21 & 0.89 & 0.50 \\
PEDD  in target per bunch                 & J/g & 0.65 & 0.40 & 0.49 & 0.66 & 1.19 \\
PEDD in target per pulse (100\,m/s)       & J/g & 61.0 & 43.7 & 41.0 & 42.4 & 45.8\\
polarization of captured positrons at DR  &  \% & 29.5 & 30.7 & 29.4 & 30.8 & 24.9 \\
\hline 
\end{tabular}
\caption{\label{tab:sourcepar}Summary of the source performance parameters for different energies and 1312 bunches per pulse. The repetition rate is 5Hz. The numbers are shown for a decelerating capture field. A flux concentrator is assumed for the studies.  See also references~\cite{ref:TDR,ref:posipol16-AU,ref:AU-250GeVthickness}.}
\end{center}
\end{table}

\section{Cooling by thermal radiation}\label{sec:radcool}
As shown in table~\ref{tab:sourcepar}, the average energy deposition is below 5\,kW assuming  nominal luminosity. In case of a luminosity upgrade, the energy deposition will be doubled. It is expected that the energy deposition in the target will be reduced by optimizing the  target thickness depending on the drive electron beam energy. Radiation cooling of few kW should be possible if the radiating surface is large enough and the heat distributes fast enough from the area of incident  beam to the large radiating surface. 
Following the Stefan-Boltzmann law, 
\begin{equation}
P = \sigma_0 \varepsilon_\mathrm{eff} A (T^4-T_0^4)\,,\label{eq:T4}
\end{equation}
with $\sigma_0 =5.67\times 10^{-12}$\,W/(cm$^2\times$ K$^4$), the effective emissivity $\varepsilon_\mathrm{eff}$ and the radiating surface $A$. 
One finds that $A=0.36\,$m$^2$ is required  to remove 2\,kW if the average temperature is $T=400^\circ$C and  $\varepsilon_\mathrm{eff}=0.5$, and $A=0.90$\,m$^2$ to remove 5\,kW. For comparison: the area of a circular ring, $r_1 = 50\,$cm, $r_2=30\,$cm is 1\,m$^2$ taking into account front and back side. 
With other words: radiative cooling is a very promising option. The wheel spinning in vacuum can radiate the heat to a stationary  cooler opposite to the wheel surface. The cooler temperature can be easily kept at room temperature by water cooling. Only in the target the heat has to be distributed fast enough from the beam path  to a larger area.

\section{Target wheel}\label{sec:wheel}
\subsection{Material parameters and temperature dependence}{\label{sec:matpar} 
The thermal conductivity of Ti6Al4V is low, $\lambda = 0.068\,$W/(cm\,K) at room temperatures and  0.126\,W/(cm\,K) at 540$^\circ$C. The heat capacity is $c=0.58\,$J/(g\,K) at room temperature and 0.126J/(g\,K) at 540$^\circ$C~\cite{ref:T-dep-parameters}. Although the wheel rotation frequency can be adjusted so that each part of the target rim is hit after 6-8 seconds, this time is not sufficient to distribute the heat load almost uniformly over a large area. For example: Following $s=\sqrt{\lambda t/\rho c}$, in Ti6Al4V the heat propagates  $s\approx 4$\,mm in 6 seconds.  The heat is accumulated in the rim and the  highest temperatures are located in a relatively small region around the beam path.  It is possible to construct a wheel consisting of the titanium alloy target rim connected with a radiator made of material with high thermal conductivity and large surface, {\it{e.g.}} with help of fins. The design optimization requires extensive FEM 
 studies 
 including the heat deposition, propagation and radiation taking into account the temperature dependent material parameters, in particular thermal conductivity, thermal capacity, density and also the emissivity of the surfaces.

For the wheel design it is important to take into account the temperature dependence of the material parameters. The Ti alloy Ti6Al4V stands high loads at operating temperatures around $400^\circ$C. At higher temperatures, special care is required to keep the the cyclic  and long-term average load below the recommended limits. Table~\ref{tab:Ti-par} summarizes important parameters and limits for Ti6Al4V used in the simulation. More details can be found in references~\cite{ref:Ti-par,ref:matweb,ref:T-dep-parameters} .

\begin{center}
\begin{table}
\renewcommand{\baselinestretch}{1.2}
\begin{tabular}{|l|c|c|cc|}
\hline
Parameter& & Unit &  & Value   \\ \hline
Density & $\rho$ &  g/cm$^3$ & & 4.43  \\
Melting point & $T_\mathrm{melt}$ & $^\circ$C &  & 1606 -- 1660  \\
Beta Transus & &$^\circ$C& & 980 \\
& & & & \\
Thermal expansion coefficient& $\alpha$ & $10^{-6}/$K & 20$^\circ$C -- 100$^\circ$C & 8,6 \\
 &  & & 20$^\circ$C -- 100$^\circ$C & 8.6  \\
 &  & & 20$^\circ$C -- 200$^\circ$C &  9.0  \\
 &  & & 20$^\circ$C -- 315$^\circ$C &  9.2  \\
 &  & & 20$^\circ$C -- 425$^\circ$C &  9.4  \\
 &  & & 20$^\circ$C -- 540$^\circ$C &  9.5  \\
 &  & & 20$^\circ$C -- 650$^\circ$C &  9.7  \\
& & & & \\
Thermal conductivity &  $\lambda$ & W/(K cm)& at RT & 0.068  \\
 &  & & at  93$^\circ$C &  0.075  \\
 &  & & at 205$^\circ$C &  0.085  \\
 &  & & at 425$^\circ$C &  0.109  \\
 &  & & at 540$^\circ$C &  0.126  \\
 &  & & at 650$^\circ$C &  0.141  \\
& & & & \\
Specific heat capacity  & $c$ & J/(g K) &at RT & 0.58  \\
 &  & & at 205$^\circ$C &  0.61  \\
 &  & & at 425$^\circ$C &  0.67  \\
 &  & & at 650$^\circ$C &  0.76  \\
 &  & & at 870$^\circ$C &  0.93  \\
& & & & \\
Spezific electric resistance &$R$& $\mu \Omega$m & RT & 1,71  \\
Modulus of elasticity &  $E$ & GPa & & 110   \\
Tensile strength &    $R_\mathrm{m}$ & MPa & &min. 828   \\
Fatigue strength (unnotched) & & MPa & &510  \\
Poisson's ratio & $\nu$ & & & 0.34 \\
\hline 
\end{tabular}
\caption{\label{tab:Ti-par}Parameters of the target material Ti6Al4V.   The values are collected from 
reference~\cite{ref:Ti-par}. }
\end{table}
\end{center}

\subsection{Wheel design options}\label{sec:w-design}
In general, two options for the target wheel  are under consideration:
\begin{enumerate}
\item
The target wheel is a  disk with the required target thickness. 
\item
The target wheel consists of a  rim made of of the target material (Ti6Al4V) which is connected to a radiator with large surface made of material with good heat conductivity. 
\end{enumerate}
In case (1) the radiating surface is limited, and due to the low thermal conductivity the gradient along the radius is large. The thermal radiation is quite low from the inner surface with low radii.  So, option (1) is only recommended for relatively low energy deposition. At elevated temperatures the material will expand. If this is prevented, the corresponding hoop stress can be roughly estimated by 
\begin{equation}
\sigma_\mathrm{hoop} = \alpha E \Delta T\,,
\end{equation}
with the thermal expansion $\alpha$, the modulus of elasticity $E$, and the temperature difference $\Delta T$.
Taking the parameters given in table~\ref{tab:Ti-par}, heating by $\Delta T = 400^\circ$C results in $\sigma_\mathrm{hoop} \approx 360\,$MPa in case that expansion is prevented. In reality, the wheel can expand. However, in the region along the beam path with highest  temperature the expansion is somewhat restricted,  hence the stress is higher (see {\it{e.g.}} reference~\cite{ref:AU-posipol14}). 
To reduce high thermal stress along the rim in the beam path area, the target could be manufactured in sectors which can expand. In a disc this can be realized by radial expansion slots. The rotation frequency has to be chosen accordingly to utilize the full available target surface. Simplicity and a relatively low weight are advantages of a disc without or with radial slots  as wheel design. The low weight is helpful  for seals and  bearing suitable for vacuum. 
\\
In case (2) the radiating surface can be easily increased by fins. The target rim needs a minimum size to cover the electromagnetic shower for the pair-production; it will have a substantially higher temperature than the radiator. To avoid high thermal stress, also the rim must be designed with sectors. Further, the connection of the target rim sectors  and the radiator must be dimensionally stable at all operating temperatures. Fins increase the mass of the radiator and have a potential for imbalances of the spinning wheel. 

For the final construction of the target wheel FEM design  studies are necessary  to ensure a long-term operation followed by systematic tests using a mock-up. 

In this paper, only option (1) is considered; the temperature and stress distribution are determined as starting point for further engineering design studies. 

\subsection{Driving mechanism and bearing}\label{sec:mech}
Magnetic bearings, developed for fly wheels for energy storage, for vacuum pumps and for Fermi Choppers have been developed, and are available on the market (SKF, Kernforschungszentrum Juelich), possibly to be adapted to the operating conditions of the rotating Titanium wheel for positron production. These are based on permanent magnet technology.
Breidenbach {\it{et al.}}~\cite{ref:breidenbach}  have studied a bearing, based on electro-magnetic coils.
Both solutions seem to be feasible. However, both of them require good radiation resistance and stability, and must be shielded. Further R\&D has therefore to be envisaged.

\section{Temperature and thermal stress distribution in the target wheel}\label{sec:Tdistr}
In case of nominal luminosity, 2\,kW are deposited in the target for  $E_\mathrm{cm}=250\,$GeV. This implies to start with a target wheel designed as disc made of  Ti6Al4V. The  temperature and thermal stress distribution are considered. 
The energy deposition is simulated with FLUKA~\cite{ref:FLUKA} and implemented into a simulation using ANSYS~\cite{ref:ansys}. Since the problem is symmetric, all results are given for one wheel sector corresponding to the length of one bunch train.
The temperature of the cooler is assumed as $T_\mathrm{cool} = 22^\circ$C. The emissivity coefficients depend on temperature and surface quality. Here, typical values given in most data sheets are used:  for  the cooler  $\varepsilon_\mathrm{cool} = 0.8$ and for the Ti6Al4V disc  $\varepsilon_\mathrm{disc} = 0.2$. An optimization of the emissivities by surface processing or coating should be possible. Such scenarios are not yet taken into account; the numbers which are given in this paper represent some kind of worst case.

The high rotation frequency of the wheel, 2000\,rpm,  yields a peak energy deposition density  of 61\,J/g pulse 
with 1312 bunches in case of the shorter distance between undulator and target (see table~\ref{tab:sourcepar}). 
The corresponding maximum temperature rise is about 105\,K. If the distance undulator to target is increased, the PEDD decreases to 43.7\,J/g, \ie{} $\Delta T_\mathrm{max} $ is nearly 80\,K. 
This temperature rise does not create shock waves since it happens within few tens microseconds.
But due to the low thermal conductivity, the temperature along the beam path region remains substantially higher than in the inner part of the wheel. In equilibrium, if the energy deposited in the wheel is equal the power radiated off the wheel, an temperature distribution is reached which can be described with an average temperature depending on the radius, $T(r)$. Only in the target wheel region around the beam path the temperature fluctuates around an average temperature with a maximum  amplitude, $T_\mathrm{max}-T_\mathrm{min}$ according to the temperature rise.
The cyclic temperature rise is much smaller than  the total temperature in the rim region. Hence, in the following subsections~\ref{sec:Tdistr-disc} and~\ref{sec:Tdistr-slot} the average temperature and stress distributions are determined since they are important to evaluate the cooling efficiency. 
 
It should be remarked that it takes a while to heat the target wheel to the equilibrium temperature where energy deposition and thermal radiation power are equal.

\subsection{Full disc target wheel}\label{sec:Tdistr-disc}
First  we considered the average temperature distribution in a wheel of 0.7\,cm thickness and derive the corresponding stress in the disc. Figure~\ref{fig:T-disc} presents the temperature for a wheel with 51 cm radius; the beam hits the target at a radius of 50\,cm.
\begin{figure}[htbp]
\center
 \includegraphics*[width=120mm]{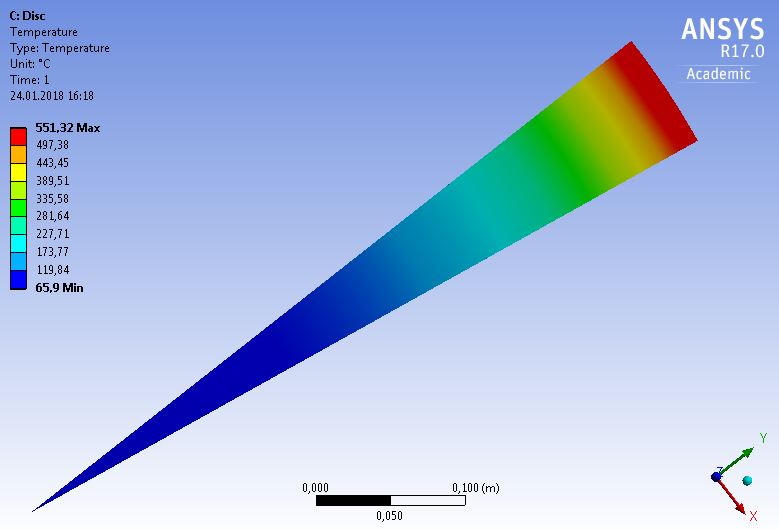}
  \caption{Temperature distribution in a target wheel consisting of a full disc with radius 51\,cm; the beam hits the target at  a radius of 50\,cm.}
 \label{fig:T-disc}
\end{figure}

 Figure~\ref{fig:disc-PS-vM}  shows  the principal stress and von-Mises stress distribution in the the target wheel. 
\begin{figure}[htbp]
\center
  \begin{tabular}{lr}
 \includegraphics[width=75mm]{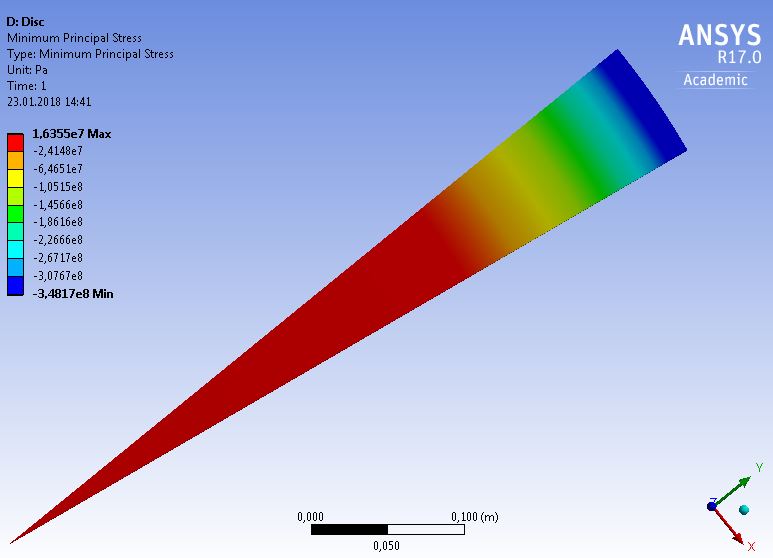}&
 \includegraphics[width=75mm]{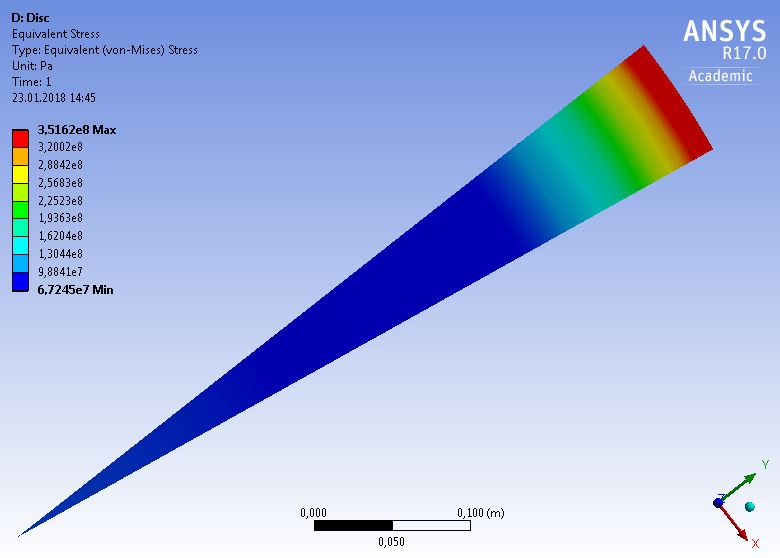}
 \end{tabular}
  \caption{Distribution of the principal stress and von Mises stress in a target wheel consisting of a full disc with radius 51\,cm; the beam hits the target at  a radius of 50\,cm.}
 \label{fig:disc-PS-vM}
\end{figure}
The maximum temperature is obtained in the region of the beam path with an average value of about $500^\circ$C. The maximum stress  in that region is about 350\,MPa. The stress distributions in figure~\ref{fig:disc-PS-vM}  represent the average.

\subsection{Full disc target wheel with radial expansion slots}\label{sec:Tdistr-slot}
A permanent stress  up to $\approx 350$MPa in the rim area is below the load limits given for Ti6Al4V. However, also the cyclic load stress caused by the pulses as well as the mechanical stress from the wheel rotation have to be taken into account. 
So it is desired to reduce this 'basic' average stress. For the time being, the influence of radial expansion slots as suggested in section~\ref{sec:w-design} has been considered for a
0.7\,cm thick disc  with 40 target sectors separated by radial expansion slots. 
In this study, the length of the slots is  20\,cm and 6\,cm. 
It is assumed that the beam hits the target at radius $50\,$cm, as radius of the target wheel $r=51\,$cm, 52\,cm and 53\,cm are considered. Since the highest temperatures are expected in the rim area, this region is most efficient in thermal radiation. So the radius of the wheel in comparison to the the beam position influences the maximum temperature in the target. 

A sketch of the target wheel with 20\,cm long radial expansion slots is shown in figure~\ref{fig:model}. 
\begin{figure}[htbp]
\center
 \includegraphics*[width=78mm]{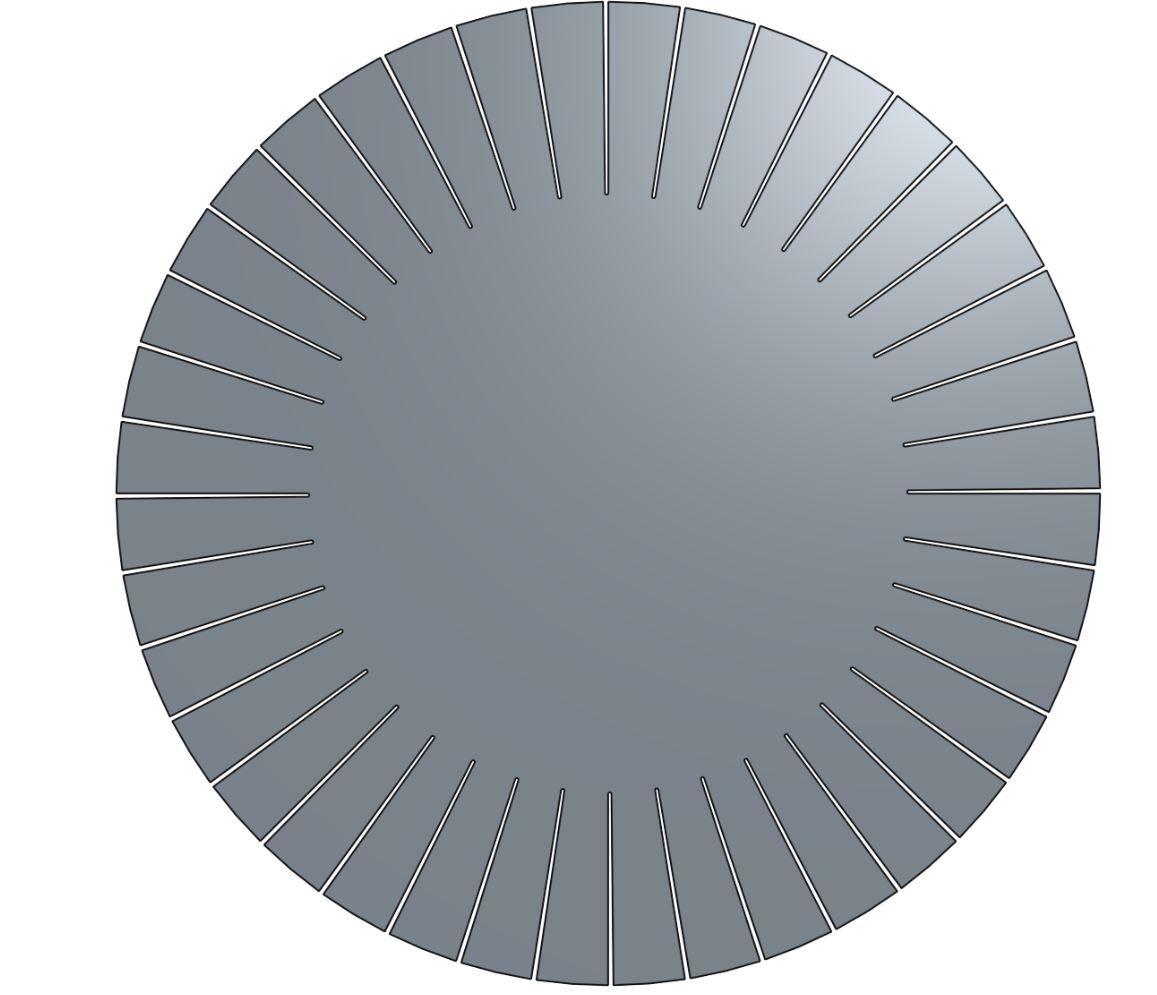}
  \caption{Sketch of the target wheel model.}
 \label{fig:model}
\end{figure}
 Figure~\ref{fig:resT} shows  the temperature distribution  and von-Mises stress for a wheel radius of $r = 52$\,cm for 20\,  long expansion slots. The expansion slots decreased the average thermal stress to very low values in the hot region around the beam path.
\begin{figure}[htbp]
\center
\begin{tabular}{lr}
 \includegraphics*[width=75mm]{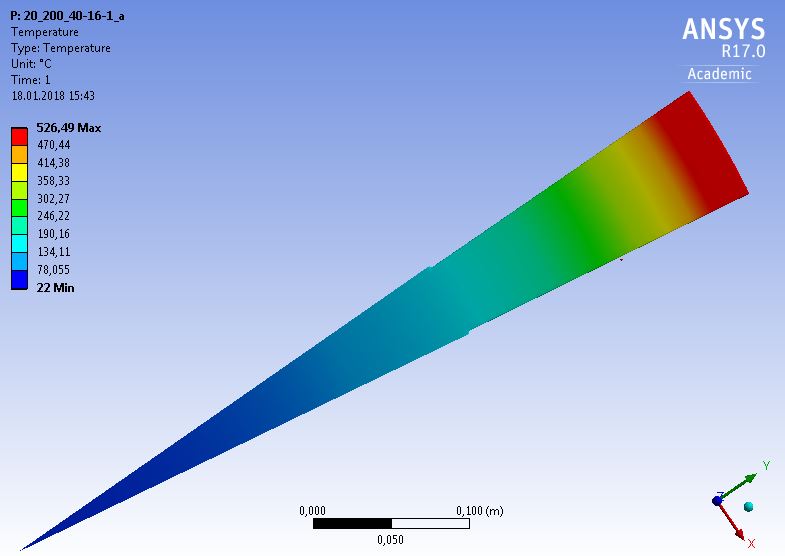}&
 \includegraphics*[width=75mm]{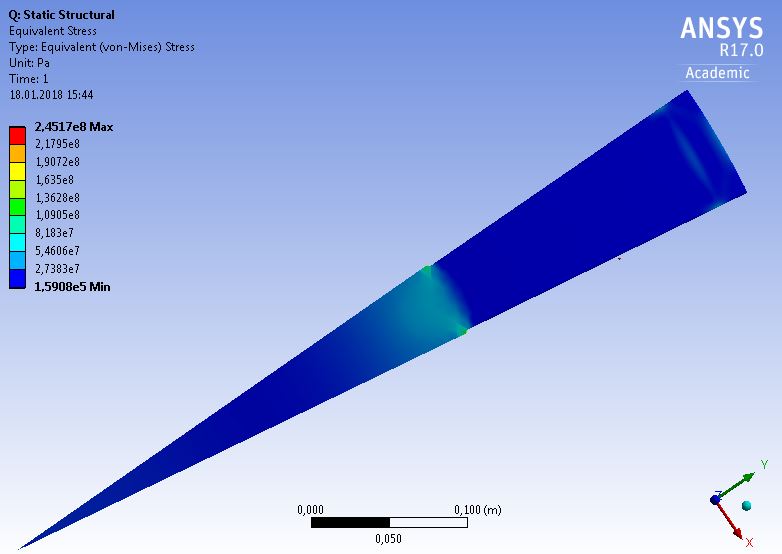}
\end{tabular}
  \caption{Temperature distribution (left) and von Mises stress distribution (right) in the target wheel with radius 52\,cm;.  the beam hits the target at  a radius of 50\,cm.}
 \label{fig:resT}
\end{figure}

Figure~\ref{fig:resTall} summarizes the radial temperature profiles for a solid disc of 51\,cm radius and a disc with expansion slots of 20\,cm length varying the wheel radius. 
\begin{figure}[htbp]
\center
 \includegraphics*[width=150mm]{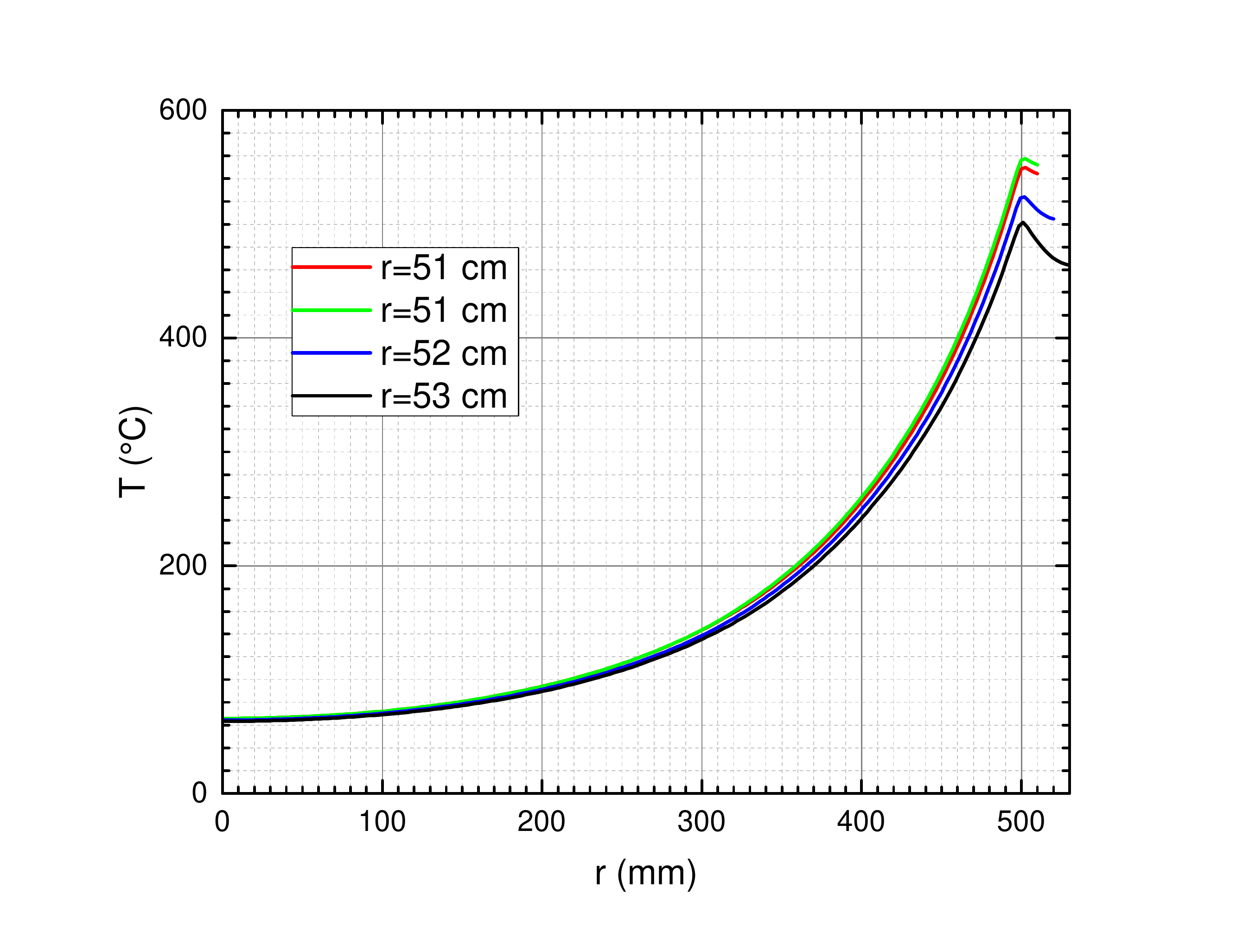}
  \caption{Radial temperature distribution in the target wheel without and with 20\,cm long radial expansion slots. Red: Disc without expansion slots, radius  51\,cm. Green: Disc with expansion slots, radius  51\,cm. Blue: Disc with expansion slots, radius  52\,cm. Black: Disc with expansion slots, radius radius 53\,cm. The beam hits the target at  a radius of 50\,cm.}
 \label{fig:resTall}
\end{figure}
As expected, highest temperatures are concentrated at the outer region of the wheel. The figure also demonstrates that a larger difference between radius of beam incidence and wheel radius decreases slightly the maximum temperature.

Figure~\ref{fig:res-PS} shows the minimal principal stress for  $r = 51; 52; 53$\,cm. 
\begin{figure}[htbp]
\center
  \begin{tabular}{lr}
 \includegraphics[width=75mm]{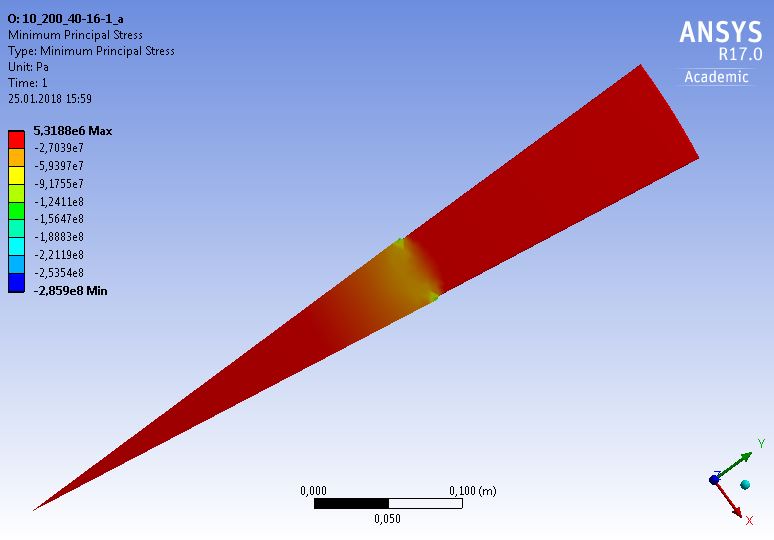}&
 \includegraphics[width=75mm]{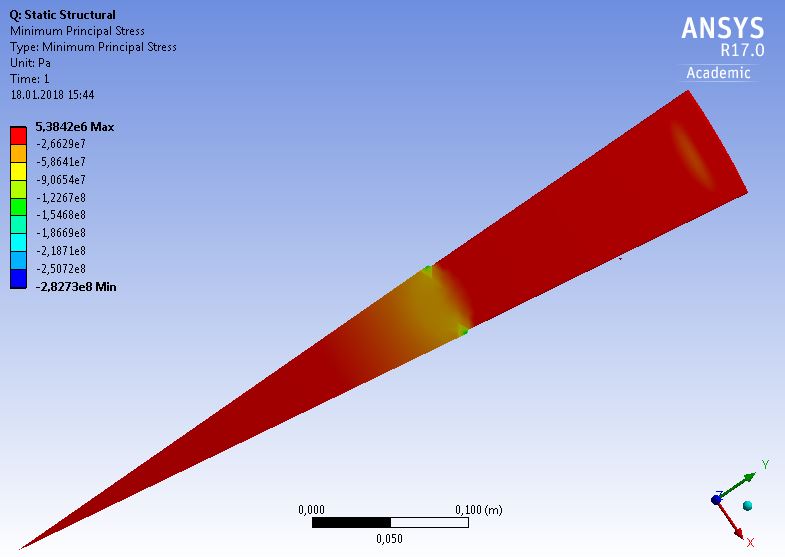}\\
 \multicolumn{2}{c}{\includegraphics[width=75mm]{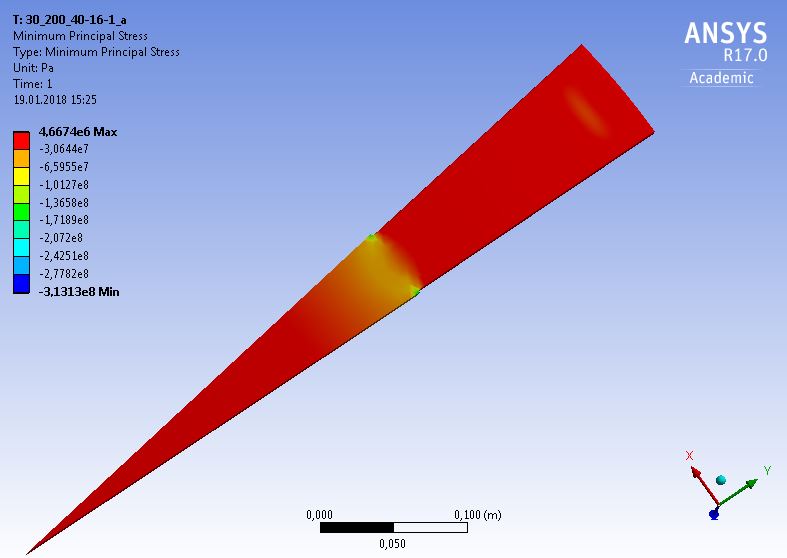}}\\
 \end{tabular}
  \caption{\label{fig:res-PS}Distribution of the minimal principal stress in the target wheel with radius 51\,cm, 52\,cm, 53\,cm; the beam hits the target at  a radius of 50\,cm. In the area of beam incidence the stress is slightly increases with the wheel radius.}
\end{figure}
The stress is small compared to that of a full wheel. Comparing the   principal stress values in the beam area,
the stress values are slightly increased for larger wheel radii since the material expansion in the most hot region is more restricted.
The same analysis was done for 6\,cm long expansion slots. The temperature distribution is almost unchanged, the von Mises stress and the minimal principal stress are shown in figure~\ref{fig:res-stress-6cm}. Indeed, expansion slots decrease substantially the stress in the beam path region but in case of too short expansion slots 
relatively high thermal stress remains along the region where the slots end.
\begin{figure}[htbp]
\center
  \begin{tabular}{lr}
 \includegraphics[width=75mm]{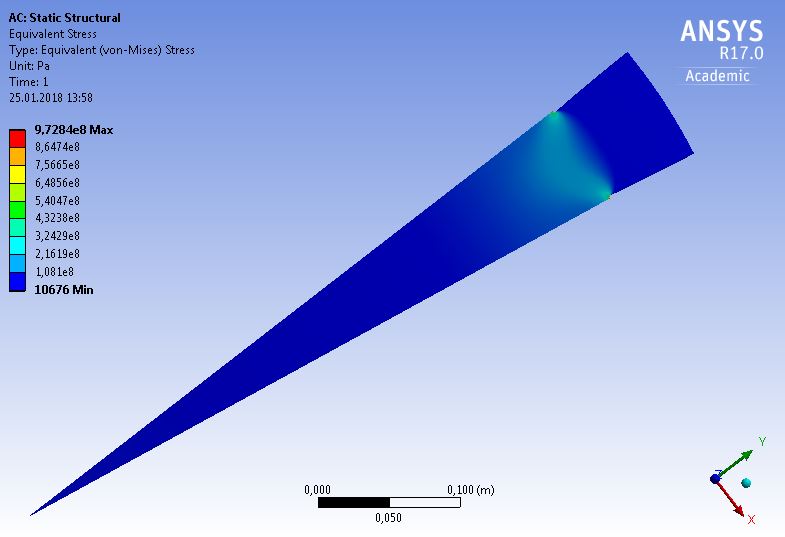}&
 \includegraphics[width=75mm]{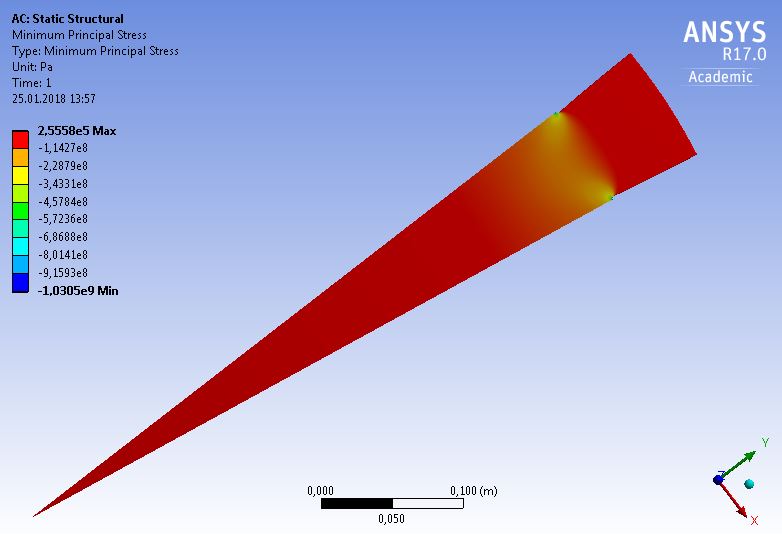}
 \end{tabular}
  \caption{Distribution of von Mises (left) and minimal principal stress (right) in the target wheel with radius 51\,cm and 6\,cm long expansion slots; the beam hits the target at  a radius of 50\,cm.}
 \label{fig:res-stress-6cm}
\end{figure}

Table~\ref{tab:stress-r} compares the  stress values at different radial positions  along the  central part of a sector of a full disc and a disc with expansion slots. Of interest are the stresses at  $r=50\,$cm (beam incidence) and at the end of the slots ($r=45\,$cm  and $r=30\,$).  
As expected, the slots reduce the thermal stress in the wheel region with slots. 

\begin{table}[hp]
\center
\renewcommand{\baselinestretch}{1.2}
\begin{tabular}{|lc|c|cc|}
\hline
& & full disc & \multicolumn{2}{c|}{expansion slots}\\
& &           &  20\,cm  & 6\,cm \\ \hline
von Mises& [MPa]&   & \multicolumn{2}{c|}{ }\\
  \multicolumn{2}{|r|}{r\,=\,50\,cm} & 348 & 7.39 & 2.71\\
  \multicolumn{2}{|r|}{r\,=\,45\,cm} & 192 & 4.43 & 125\\
  \multicolumn{2}{|r|}{r\,=\,30\,cm} & 67.66 & 47.4 & 44.5\\\hline
min. principle &[MPa] &   &\multicolumn{2}{c|}{ }\\
  \multicolumn{2}{|r|}{r\,=\,50\,cm} & -347  & -8.47 & -3.59\\
  \multicolumn{2}{|r|}{r\,=\,45\,cm} & -172   & -0.005 & -110\\
  \multicolumn{2}{|r|}{r\,=\,30\,cm} & 0.0002 & -42.08  & -4.75\\\hline
\end{tabular}
\caption{\label{tab:stress-r}Von Mises and minimal principal stress for
different radii along the central part  of a sector. The wheel radius is
51\,cm. The minus sign corresponds to compressive stress.}
\end{table}

\subsection{Stress at radial expansion slots}\label{sec:slot-end}
A closer look  reveals  high stress values  in a  small region at the end of the expansion slots  as shown in figure~\ref{fig:dehnung}. 
For example: For wheel radii
 of 51\,cm, 52\,cm and 53\,cm simulations were performed with 20\,cm long expansion slots. Corresponding to the wheel temperature, the stress increases slightly in the region at the end of the slots with highest values in the arc. 
But all values are still below the material load limit at this temperature. 
\begin{figure}[htbp]
\center
  \begin{tabular}{lr}
 \includegraphics[width=75mm]{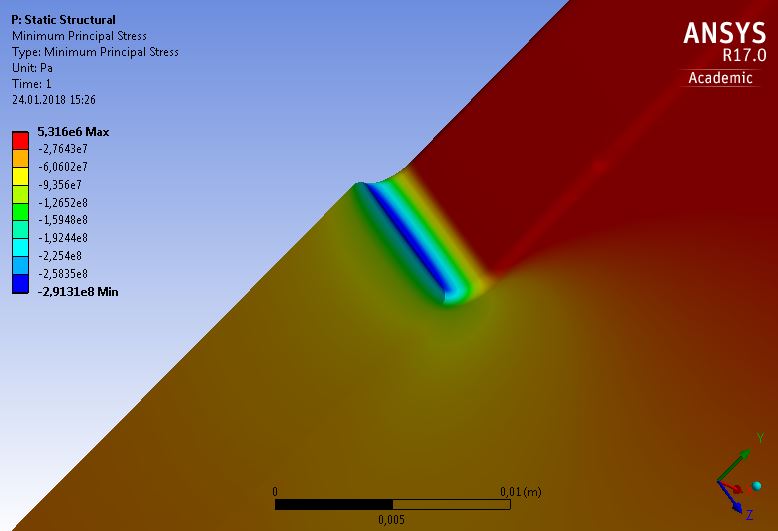}&
 \includegraphics[width=70mm,height=50mm]{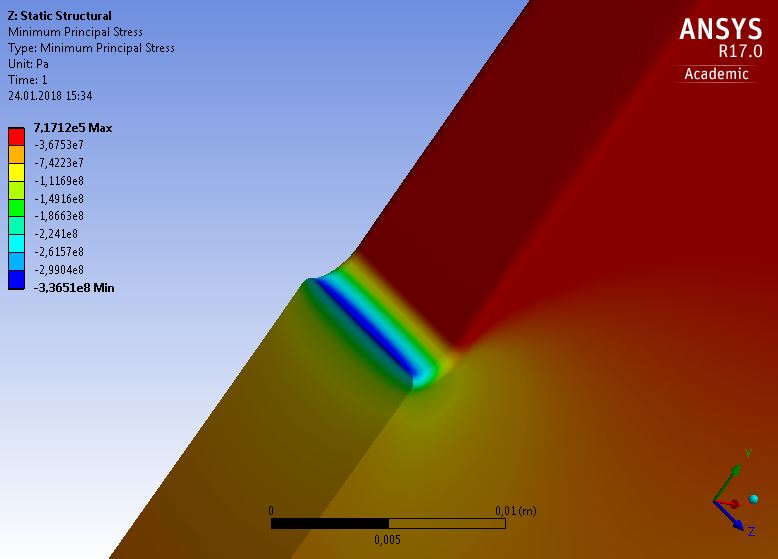}
 \end{tabular}
  \caption{Distribution of the principal stress at the end of the 20\,cm long radial expansion slots the target wheel. The wheel radius is 51\,cm (left), and 53\,cm (right); the beam hits the target at  a radius of 50\,cm.}
 \label{fig:dehnung}
\end{figure}

For comparison, also the thermal stress in 6\,cm long expansion slots  was studied. The temperature distribution remained unchanged, but the stress increased substantially in the region circumferential along the end of the slots.  In particular, in the arc region of the slots stress values beyond the material limits are obtained.  The results are shown in figure~\ref{fig:expslot-short}.
\begin{figure}[htbp]
\center
  \begin{tabular}{lr}
 \includegraphics[width=75mm]{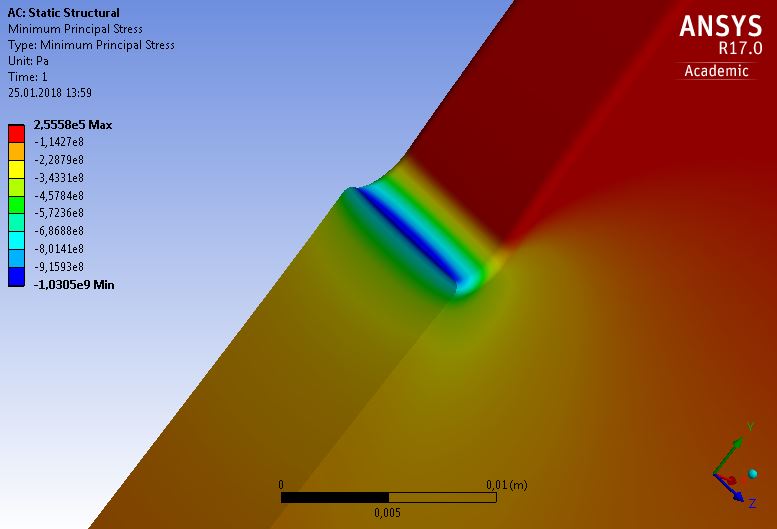}&
 \includegraphics[width=75mm]{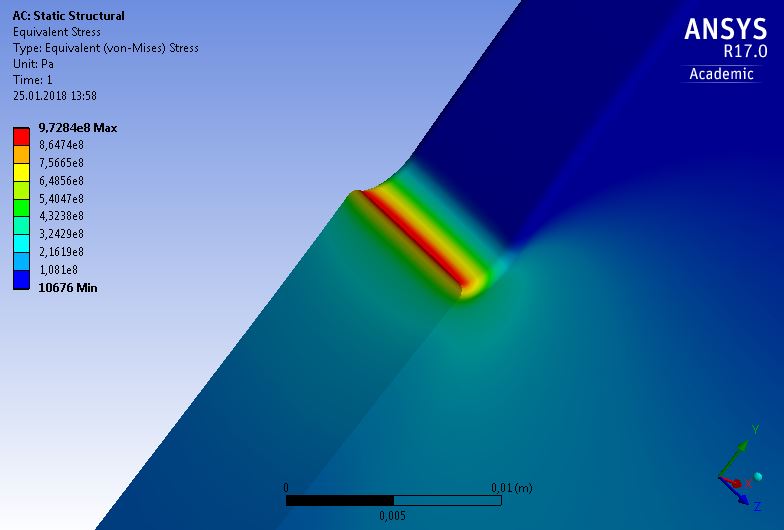}\\
 \end{tabular}
  \caption{Distribution of the minimal principal stress and von Mises stress at the end of the 6\,cm long radial expansion slots. The wheel radius is 51\,cm; the beam hits the target at  a radius of 50\,cm.}
 \label{fig:expslot-short}
\end{figure}

It is expected that the shape of the slot ends  influences the stress significantly. An optimized shape 
will reduce the stress values in the arcs, for example  by drilling a larger whole at the slot end. However, the length of the expansion slots is essential. 

\subsection{Cyclic load}\label{sec:cyclic}
Ti6Al4V offers excellent thermal and mechanical properties up to temperatures of about 400--500$^\circ$C. However,  the longterm stability under cyclic load depends on the temperature and material conditions and  has to be tested for the concrete application.
The photon beam creates cyclic load in the ILC positron target which happens every 6-8 seconds at the same position (depending on the wheel revolution frequency).
The maximum instantaneous temperature rise per pulse depends on the PEDD. Taking the numbers given in table~\ref{tab:sourcepar} one finds a maximum temperature rise  of about 80\,K  for electron beam energies $E_\mathrm{e^-} \ge 150\,$GeV if a flux concentrator is used and the target is at room temperature. For the low energy case,  $E_\mathrm{e^-} \approx 125\,$GeV, the lower drift space yields a higher peak energy deposition density of 61\,J/g 
corresponding to maximum temperature rise per pulse of 105\,K in a target at room temperature. However, the heat capacity increases with temperature (see table~\ref{tab:Ti-par}) so that the temperature rise is lower for high target temperatures, {\it e.g.} about 90\,K instead of 105\,K in case the target is heated to the equilibrium temperature.   
Assuming, that in   the very short time  of energy deposition the heated volume cannot expand it 
undergoes a stress of
\begin{equation}
\sigma = \frac{\alpha E \Delta T}{1-2\nu}\,, \label{eq:th-stress-volume}
\end{equation} 
where $\nu$ is the Poisson ratio. This cyclic stress occurs in addition to the thermal stress in the heated target. 
With equation~\ref{eq:th-stress-volume} one finds maximum values for the cyclic stress  of about 260\,MPa per pulse for $\Delta T_\mathrm{max} = 80\,$K and about 340\,MPa for $\Delta T_\mathrm{max} = 105\,$K assuming an instantaneous temperature rise. 
Since the target is moving with 100\,m/s, for an electron beam of 126.5\,GeV almost 100 of the 1312 bunches of one pulse are superposed  corresponding to a load time interval of about 55 microseconds. The target material is heated  along the beam path corresponding to the electromagnetic shower, {\ie} the heated volume can expand. In that sense, the cyclic peak stress values will be substantially below the results calculated with equation~\ref{eq:th-stress-volume}. 

The cyclic peak stress values are below the fatigue limit given in table~\ref{tab:Ti-par}. In addition, the cyclic stress is only local and the peaks occur only in small regions.  
Studies have shown that stress waves are not relevant since the time of load impact is relatively long. 

It is important to keep the  thermal stress in the wheel at a level that the total stress --including the cyclic load-- does not exceed the stress limit, for example with help of expansion slots. 
But with  expansion slots the beam impact has to be synchronized with the target sectors. In addition to the very tight rotation frequency control also a very precise position control of the wheel is required to ensure a constant  beam intensity during the full pulse. 

Ignoring the stress created by the wheel rotation, a maximum equivalent stress  of 348\,MPa is found in the beam path region of a full disc wheel (see table~\ref{tab:stress-r}). Adding the cyclic stress caused by the pulsed photon beam,  the stress would reach peak values which most likely exceed the load limit for long-term operation at elevated temperatures.  
But if the maximum temperature rise per pulse could be reduced, for example, by a longer drift space which reduces the PEDD,  even a wheel without expansion slots could be an option.


The pulsed heating of parts along the target rim could yield imbalances. First studies~\cite{ref:FS-imbalances} showed that they are expected to be small. Further studies will follow. 
  
\subsection{Target material test}\label{sec:mami}
The fatigue load limit depends on the temperature and number of load cycles; corresponding numbers can be found partially in tables with material properties. But in addition to thermal and mechanical stress the target is passed by a photon  beam which could change the material properties by damaging the material structure. To study this, experimental tests were performed.   
 To simulate a load corresponding to that expected at the ILC positron target, the 14\,MeV electron beam of the Mainz Mictrotron (MAMI) injector was used.  
It was managed to focus the the rms spot size on target to $\sigma \approx 0.2\,$mm for  the chopped 50\,$\mu$A cw electron beam with pulse length of 2\,ms~\cite{ref:mami-beam}.   With a repetition rate of 100\,Hz up to $6.6\times 10^6$ load cycles were generated. This number of cycles will be reached after at least 2 years ILC running time.  All irradiated samples survived the irradiation procedure without damage visible by eyes. The grain structure was modified in samples that  reached maximum temperatures near to the phase transition value. For the details see reference~\cite{ref:MAMI-IPAC17}. It was concluded that the operation of a positron target consisting of Ti6Al4V is possible if the maximum temperature corresponds to the recommended operation temperature and exceeds only locally for short time this level up to  about $700^\circ$C. Further studies are planned to test and confirm the robustness of target materials~\cite{ref:gudi-bmbf}.
Very high temperature Titanium alloys, for example Ti SF-61 could be an alternative as suggested in reference~\cite{ref:breidenbach}; SF-61 stands higher operation temperatures than Ti6Al4V. 

As already mentioned, the heating of the target yields a non-uniform temperature distribution and stress within a wheel. At a first glance~\cite{ref:FS-imbalances}, the corresponding deformation due to expansion does not yield imbalances of the spinning wheel.  
So far, the dynamic effects have not yet considered in detail. Comprehensive simulations are planned to study them in order to prepare a reliable wheel design.

\subsection{Target wheel and OMD}\label{sec:T+omd}
Following the ILC TDR, the target wheel will be installed near to a pulsed flux concentrator (FC). Alternatively, a quarter wave transformer (QWT) can be used to capture and focus the positrons. In both cases the magnetic field in the target matters since eddy currents could increase the target temperature and could slightly drag the wheel rotation. First calculations of eddy currents as well as tests with a Ti6Al4V wheel spinning in a  magnetic field have shown that eddy currents are not a serious problem for the target operation~\cite{ref:sievers-eddy,ref:UK-eddy}.  However, the distances between  target and flux concentrator or QWT must be very small to achieve a high positron yield. Both, FC and QWT occupy a large area in front of the spinning target since the radii  of these devices are about $25-30$\,cm.  Passing the area of these matching devices the hot target surface cannot radiate to the cooler surfaces. The results presented in section~\ref{sec:Tdistr} do not take in
 to account the somewhat lower cooling efficiency due to QWT or FC. Either FC or QWT can be cooled excellently or the target temperatures are slightly higher. More important is the energy which is deposited by the particle shower in  FC and QWT. The positron capture efficiency and the aperture of the matching device are still under study~\cite{ref:AU-OMD}.

\section{Summary}\label{sec:sum}
First studies for  $E_\mathrm{cm}=250$\,GeV are performed  to design a positron target for the undulator based positron  source cooled by thermal radiation. Such design is expected to be substantially more easy than a target wheel with water cooling. 
Assuming nominal luminosity, an average power of 2\,kW is deposited in a 0.2\,$X_0$ thick Ti6Al4V target. However, depending on the detailed positron source layout, \eg{} drift space between undulator and target, the peak energy density  deposited per pulse in the target (PEDD) could be high. 
The results shown in this paper correspond to a source parameter option with highest load per pulse in the target. The material parameters for target and cooler were not optimized; for example, the cooling efficiency could be substantially improved by optimizing the surface emissivities, also titanium alloys with higher heat resistance should be considered. 
No constraints have been identified which  preclude radiation cooling.  The studies show that a rotating disc with radial expansion slots would  be an option. Even a simple full disc could be a useful design if the PEDD  is lower than that considered in this paper (see table~\ref{tab:sourcepar}).  
More studies and engineering work is required to approach a final target wheel design; 
in particular, dynamic stress  will  be taken into account.

\section*{Acknowledgment}
This work was supported by the German Federal Ministry of Education and Research,
Joint Research Project R\&D Accelerator ``Positron Sources'', Contract Number 05H15GURBA.

\end{document}